\documentclass{article}
\usepackage{spconf,amsmath,graphicx}
\usepackage{amssymb,amsfonts,bm}

\usepackage{enumitem}
\setlist{nosep, leftmargin=14pt}

\usepackage{mwe} 
\usepackage{xcolor}
\usepackage{hyperref}

\title{Cross-level Contrastive learning and consistency constraint for semi-supervised medical image segmentation}
\name{Xinkai Zhao$^1$, Chaowei Fang$^{2*}$, De-Jun Fan$^3$, Xutao Lin$^3$, Feng Gao$^3$, Guanbin Li$^{1*}$
\thanks{$^*$ Corresponding Authors.}}
\address{$^1$School of Computer Science and Engineering, Sun Yat-Sen University, Guangzhou, China\\ 
$^2$School of Artifical Intelligence, Xidian University, Xi'an, China \\
$^3$The Sixth Affiliated Hospital, Sun Yat-sen University, Guangzhou, China }

\begin{document}
%
\maketitle

\begin{abstract}

Semi-supervised learning (SSL), which aims at leveraging a few labeled images and a large number of unlabeled images for network training, is beneficial for relieving the burden of data annotation in medical image segmentation. 
According to the experience of medical imaging experts, local attributes such as texture, luster and smoothness are very important factors for identifying target objects like lesions and polyps in medical images.
Motivated by this, we propose a cross-level contrastive learning scheme to enhance representation capacity for local features in semi-supervised medical image segmentation.
Compared to existing image-wise, patch-wise and point-wise contrastive learning algorithms, our devised method is capable of exploring more complex similarity cues, namely the relational characteristics between global and local patch-wise representations.  
Additionally,  for fully making use of cross-level semantic relations, we devise a novel consistency constraint that compares the predictions of patches against those of the full image.
With the help of the cross-level contrastive learning and consistency constraint, the unlabelled data can be effectively explored to improve segmentation performance on two medical image datasets for polyp and skin lesion segmentation respectively.
Code of our approach is available 
\href{https://github.com/ShinkaiZ/CLCC-semi}{\color{blue}{here.}}

\end{abstract}
\section{Introduction}
\label{sec:intro}

The optimization of segmentation models based on convolutional neural networks (CNN) is very data-hungry, which requires a large number of carefully annotated images.
Semi-supervised learning, which requires limited amount of annotated data and large amount of unlabeled data, can help to reduce the workload of annotation.
In this topic, the reasonable exploration of unlabeled data is critical for preventing the segmentation models from overfitting with scarce annotated data.

A classical method for involving in unlabeled images is the self-labeling algorithm~\cite{lee2013pseudo}, which attempt to automatically allocate pseudo labels for unlabeled images with self-learned models. The performance of this kind of methods is dependent to the quality of pseudo labels. Besides, they may be  severely influenced by the confirmation bias since the self-labeling strategy is difficult to rectify the incorrect predictions.    
The other typical manner for taking advantage of unlabeled images is the  self-supervised learning~\cite{doersch2015unsupervised} which is effective visual representations from unlabelled data with upstream pretext tasks. 
These pretext tasks such as relative position prediction~\cite{doersch2015unsupervised} and rotation prediction~\cite{gidaris2018unsupervised} usually can be easily annotated and have underlying relations to the understanding of images.
They can be exerted to enhancing the representation capacity of segmentation models.
Owing to the high interpretability, contrastive learning attracts extensive research interest and brings about significant progress in self-supervised learning. 

\begin{figure}[tb]
	\centering
	\includegraphics[width=0.48\textwidth]{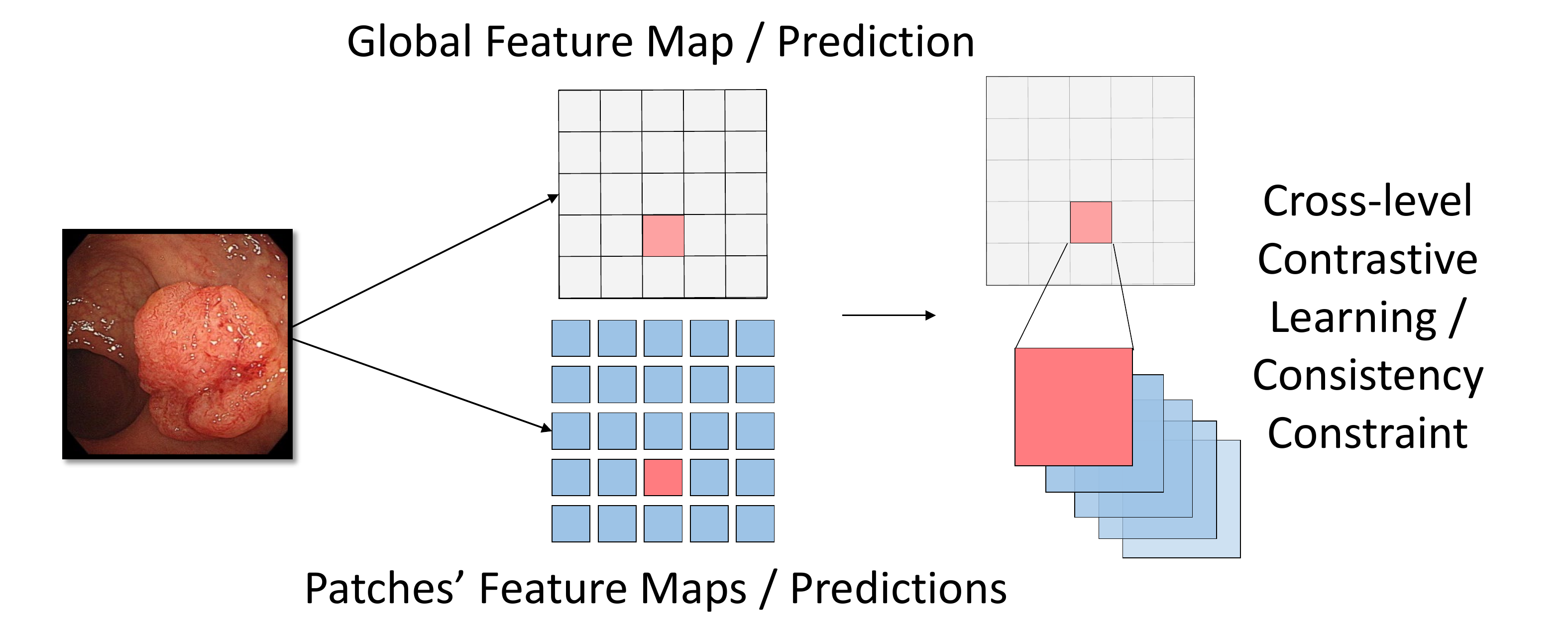}
\caption{The motivation and the core idea of this paper. We apply contrastive learning or consistency constraint between features or predictions obtained from the full image and local patches, thus encouraging the network to learn the local characteristics in semi-supervised medical image segmentation.}
	\label{fig:example}
 	\vspace{-3mm}
\end{figure}

Contrastive learning (CL) algorithms~\cite{hadsell2006dimensionality,chen2020simple,he2020momentum} usually regard similar image pairs as positive instances and dissimilar image pairs as negative instances. 
During training, positive instances are encouraged to get closer to each other while negative instances are constrained to be mutually exclusive in the latent feature space. 
The majority of CL approaches in image classification tasks follow the pipeline as SimCLR~\cite{chen2020simple}, by applying contrastive metric learning constraints on image-level embeddings.
However, for pixel-level dense prediction tasks like image segmentation, such image-level CL approaches ignore the spatial relation information of the image. 
For exploring spatial relation information, \cite{xie2021propagate} and~\cite{lai2021semi} construct positive instances with pairs of points under different augmentations but having the same actual locations, and select the negative instances according to semantic inconsistency.
\cite{wang2021dense} combines both image-level and point-level CL, and relies on the correspondence between different views of the same image to sample positive instances for the point-level CL.
Similarly, \cite{chaitanya2020contrastive} exploits both image-level and point-level CL to enhance the representation capacity for semi-supervised medical image segmentation. 
However, as far as we know, existing CL methods merely compare features within the same level, namely constraining image-level, patch-level or point-level features.
The more complicate cross-level similarity cues remains under-explored.

In medical image analysis, local characteristics such as textures and smoothness are critical factors for the identification of the target objects like polyps, tumors and organs. 
These features are gradually diluted as the forward propagation reaches deeper layers in CNN models. 
Aiming to strengthen the representation capacity on these local features, we propose a cross-level contrastive learning scheme for semi-supervised medical image segmentation. Provided a full image, a U-Net~\cite{ronneberger2015u} is employed for extracting the global dense featuremap and deriving of dense predictions.
For constraining the receptive field within a limited space, the full image is decomposed into small patches which are successively fed into the CNN model for generating the local dense featuremap. 
Afterwards, for each point in the global dense featuremap, the point with the same location in the local dense featuremap is regarded as the positive sample, and a set of points at other locations are randomly selected from the global dense featuremap as negative samples. Our proposed cross-level contrastive learning scheme is implemented based on the above tuple formulation strategy.
Furthermore, for fully taking advantage of the cross-level relation knowledge from the perspective of semantic content, we apply a consistency constraint between dense predictions of the full image and local patches.
The semantic consistency between the two prediction levels can  help to promote the intra-class compactness for both global and local dense features.
We evaluated our approach on two publicly available medical image segmentation datasets including  a polyp segmentation dataset (Kvasir-SEG~\cite{jha2020kvasir}) and a skin lesion segmentation dataset (ISIC 2018~\cite{codella2019skin}). Elaborate experiments show that our approach outperforms existing semi-supervised medical image segmentation methods significantly.
Main contributions of this paper are summarized as follows.
\begin{itemize}
    \item [1)] We devise a cross-level contrastive learning algorithm to enhance the representation capacity for local features in semi-supervised semantic segmentation.
    \item [2)] A cross-level consistency constraint is devised to transmit the representational similarity to the final prediction and explore the semantic relation across levels.
    \item [3)] Experiments on poly and skin lesion segmentation tasks indicate that our proposed method derives of evidently better performance that state-of-the-art semi-supervised segmentation methods. 
\end{itemize}

\begin{figure*}[t]
	\centering
	\includegraphics[width=0.9\textwidth]{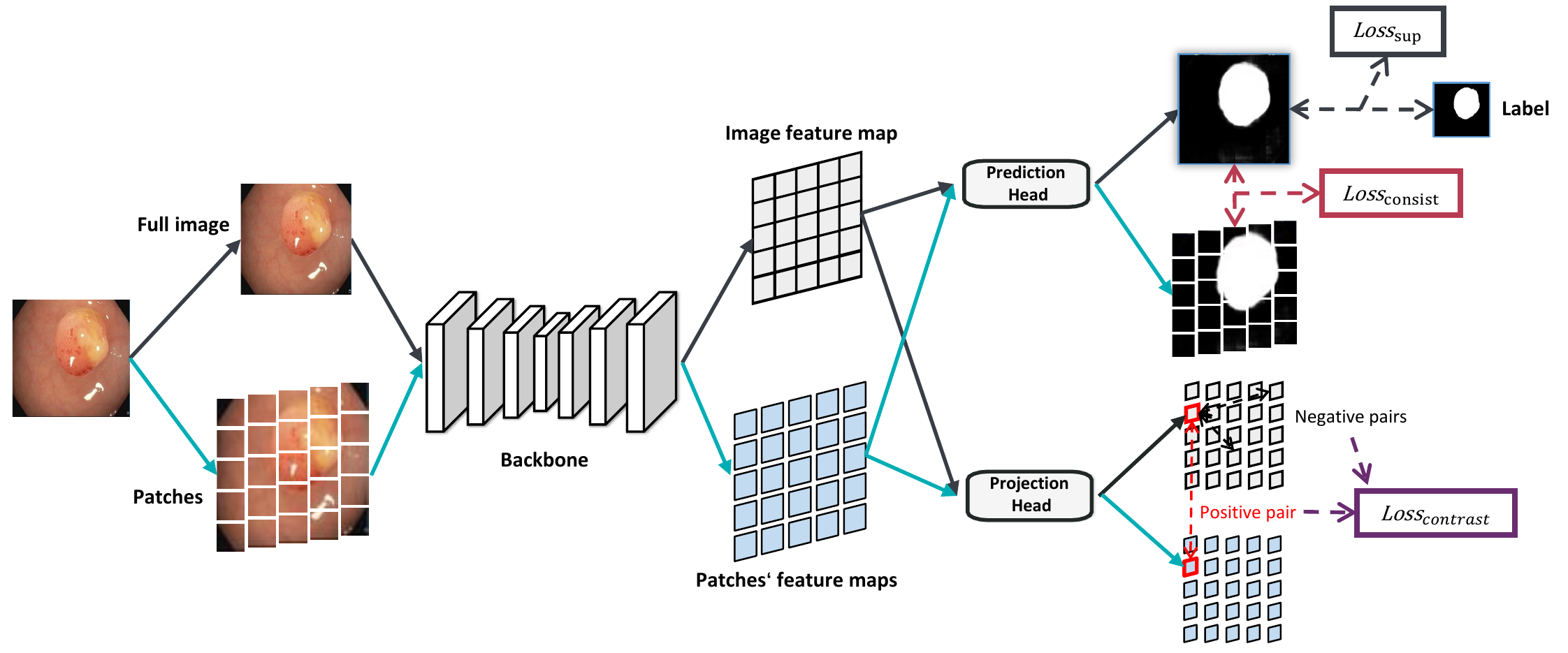}
	\vspace{-3mm}
	\caption{The overview of our approach for semi-supervised medical image segmentation.}
	\label{fig:framework}
	\vspace{-2mm}
\end{figure*}

\section{Method}
\label{sec:format}

This paper is targeted at tackling the semi-supervised medical image segmentation task. Suppose we have a medical image dataset which is constituted by $N$ images $X_L = \{\mathcal{X}_i\}_{i=1}^N$ with pixel-wise annotations $Y_L = \{\mathcal{Y}_i\}_{i=1}^N$, and $M$ images $X_U = \{\mathcal{X}_i\}_{i=N+1}^{N+M}$ without annotations. 
The goal is to learn a model with strong generalization capacity on unseen testing images.
The framework of our approach is shown in Fig.~\ref{fig:framework}, which is composed of three losses including a supervised loss, a dense cross-level contrastive loss, and a dense cross-level consistency constraint.

\subsection{Dense Cross-level Contrastive Learning}
\label{sec:cl}

Different with many existing works that extract representations just in image level or patch level, we propose a novel dense cross-level contrastive loss.
In our approach, backbone takes both global image and local patches as input,  extracting features to compare,
thus sufficiently exploring the representational information of respective details in the image.

For input image $\mathcal{X} \in \mathbb{R}^{H*W*3}$, we use a U-Net structure backbone $f$ to extract the feature map $\mathcal{Z}=f(\mathcal{X})$.
Meanwhile, for each input image $\mathcal{X}$, we decompose one image into $n*n$ patches $ {x}_i \in \mathbb{R}^{\frac{H}{n}*\frac{W}{n}*3}, i \in [1...n*n]$.
Thereafter, we use the same backbone to extract a local feature map for each patch, 
${z}_i=f({x}_i)$.
Both global feature map $\mathcal{Z}$ and local feature maps $\{{z}_i\}$
are fed into a same projection head $p$, constituted by three 
convolutional layers, resulting to features $p(\mathcal{Z}) \in \mathbb{R}^{n*n*D}$ and $p({z}_i) \in \mathbb{R}^{1*1*D}$.

We expect that the representations obtained from patches are similar to those derived from the entire image.
Therefore, we design a patch-image contrastive learning loss.
Specifically, we utilize a projection head to project feature $\mathcal{Z}$ to $p(\mathcal{Z})$, where the perceptual fields of each pixel in $p(\mathcal{Z})$ does not overlap on $\mathcal{Z}$.
The feature map of the entire image $p(\mathcal{Z})$ is decomposed into $n$*$n$ blocks, denoted as $p(\mathcal{Z})_i \in \mathbb{R}^{1*1*D}, i \in [1...n*n]$.
For the two feature sets, $\{p({z}_i)\}$ and $\{p(\mathcal{Z})_i\}$, we use the following formula to calculate the contrastive loss:
\begin{align}
\small
    & \mathcal{L}_\text{contrast}(i) = \notag \\
    &-\log \frac{\exp \left[p(\mathcal{Z})_i \cdot p({z}_i) / \tau\right]}{\exp \left[p(\mathcal{Z})_i \cdot p({z}_i) / \tau\right]+\sum \limits_{n \neq i} \exp \left[p(\mathcal{Z})_i \cdot p(\mathcal{Z})_n / \tau\right]}
\end{align}
where $p(\mathcal{Z})_i$ is a pixel of the projected feature map $p(\mathcal{Z})$;
$p(\mathcal{Z})_i$ and $p({z}_i)$ are the feature vectors in the same position $i$, and are regarded as the positive pair of contrastive learning; 
$\tau$ is a scalar temperature hyperparameter. 
The loss is averaged over all pixels in the feature map.

\subsection{Dense Patch-image Consistency Learning}
\label{sec:methodconsistency}

Contrastive learning is able to enhance the similarity between patches and image at the feature level. In order to transmit the representational similarity to the final prediction and further enhance the context-independent prediction ability of the network, we apply a unsupervised consistency loss between the complete prediction and partial predictions of an image.
Specifically,
for global feature map $\mathcal{Z}$ and local feature maps $\{{z}_i\}$ of the same image, we utilize a prediction head $h$, a single convolutional layer with the convolutional kernel of 1*1., to generate the segmentation predictions.
We use mean-square error to measure the difference between the predicted results:
\begin{align}
\vspace{-4mm}
\small
     \mathcal{L}_\text{consist} = \sum \limits_{i} MSE(h(z_i), h(\mathcal{Z})_i)
\vspace{-4mm}
\end{align}
where $h(\mathcal{Z})_i$ means the $i$-th patch having the same position as $z_i$ in the prediction result of the entire image.

\subsection{Overall}
\label{sec:overall}
The overall loss function is formed by summing the supervised loss and the unsupervised loss.
\begin{align}
\small
    \mathcal{L}_{all} = \mathcal{L}_{sup} + \alpha \mathcal{L}_{contrast} + \beta \mathcal{L}_{consist}
\end{align}
For the supervised loss on labeled images, we adopt the combination of a Dice loss and a cross entropy loss:
\begin{align}
\small
    \mathcal{L}_{sup} = \frac{1}{2} {\left(
    {DICE} \left( h(\mathcal{Z}), \mathcal{Y} \right) + {CE} \left( h(\mathcal{Z}), \mathcal{Y} \right)
    \right)}
\end{align}

The practical training process is consisting of two stages.
In the first stage, we use the supervised and contrastive loss to train the network backbone, aiming to enable the network to extract generalized representations. Namely, $\alpha$ and $\beta$ is set to 1 and 0 respectively.
In the second stage, we use consistency loss to 
further utilize cross-level relations in the image and 
strengthen the network's ability of capturing useful partial information. 
Namely, $\alpha$ and $\beta$ is set to 0 and 1 respectively.
The overall training process has 300 epochs, including 100 epochs for the first stage and 200 epochs for the second stage.

\section{Experiments}
\label{sec:exp}

We evaluate our proposed semi-supervised learning approach on two medical segmentation datasets. Comparisons with existing state-of-the-art methods including
UAMT \cite{yu2019uncertainty}
and
URPC \cite{luo2020efficient}
are conducted to showcase the efficacy of our method.

\subsection{Datasets}
\label{sec:datasets}

\textbf{Kvasir-SEG dataset} \cite{jha2020kvasir} is a polyp segmentation dataset, which contains 1,000 white-light colonoscopy images with pixel-level manual labels. We randomly choose 60\% images as the training set, 20\% images as the validation set, and the remaining 20\% images as the testing set. 
Among the training set, we select 20\% (120) images as labeled data and the other 80\% (480) as unlabeled data.

\textbf{ISIC 2018 dataset} \cite{codella2019skin} is a skin lesion segmentation dataset.
We use 2594 skin images with pixel-level lesion labels. 
We randomly choose 60\% images as the training set, 20\% images as the validation set, and the remaining 20\% images as the testing set. 
Among the training set, we select 10\% (156) images as labeled data and the other 90\% (1400) as unlabeled data.

\subsection{Implementation Details and Evaluation Metric}
\label{sec:details}
Our approach is implemented in Python with PyTorch library. 
We used NVIDIA Tesla A100 GPUs for training and testing. 
The network is trained using Adamw~\cite{loshchilov2017decoupled} optimizer with default parameters setting. 
Every training mini-batch is consisting of four annotated images and four unannotated images. 
All images are resized into 320*320.
For fair comparison,  the same backbone model, supervised training loss, dataset division and epochs are used 
for all methods. 

We use mean absolute error (MAE), mean intersection over union (mIoU) and Dice coefficient of foreground to evaluate the performance of each method.
For each experiment, we randomly select labeled images for 3 times and report the mean and standard deviation.

\vspace{-2mm}
\subsection{Results}
\label{sec:results}
\begin{table}[t]
 \footnotesize
\caption{Comparison of our approach with state-of-the-art semi-supervised medical image segmentation approaches on Kvasir-SEG dataset with 120 labeled images and 480 unlabeled images.}
\label{tab:tab1}
\begin{center}
\begin{tabular}{l | c c c  }
\hline
Method  &MAE &Dice &mIoU\\
\hline
U-Net    &$8.16 $ &$67.03 \pm 0.87$ &$57.08 \pm 0.64$ \\
UAMT    &$7.40 $ &$69.87 \pm 0.61$ &$59.43 \pm 1.11$ \\
URPC    &$7.10 $ &$70.83 \pm 0.24$ &$61.50 \pm 0.71$ \\
\hline
ours (w/o consist)  &{$7.73$} &{$70.37\pm0.94$} &{$59.73\pm0.97$} \\
ours (w/o contrast)  &{$7.77$} &{$72.17\pm1.46$} &{$61.10\pm1.56$} \\
ours (all)  &\bm{$6.63$} &\bm{$73.63\pm0.25$} &\bm{$63.50\pm0.16$} \\
\hline
\end{tabular}
\vspace{-6mm}
\end{center}
\end{table}


Table \ref{tab:tab1} shows the results of the semi-supervised segmentation algorithms on Kvasir-SEG dataset. 
Compared with the baseline and other existing SOTA methods, our approach achieves a significant and consistent improvement. 
In addition, if we just use consistency loss without contrstive loss,
or use contrstive loss without consistency loss, all the evaluation metrics are improved but not stable.
The result of the ablation experiments demonstrates that both of our proposed loss functions can significantly improve the performance of the network. 
Table \ref{tab:tab2} shows the results in ISIC 2018 dataset. Our approach also shows consistent effectiveness on ISIC 2018 dataset.

\begin{table}[t]
 \footnotesize
\caption{Comparison of our approach with state-of-the-art methods on ISIC 2018 dataset with 156 labeled images and 1400 unlabeled images.}
\label{tab:tab2}
\begin{center}
\begin{tabular}{l | c c c  }
\hline
Method  &MAE &Dice &mIoU\\
\hline
U-Net    &$8.03 \pm 0.98$ &$79.2 \pm 0.93$ &$70.77 \pm 1.16$ \\
UAMT    &$7.13 \pm 0.94$ &$80.70 \pm 1.87$ &$72.16 \pm 2.10$ \\
URPC    &$7.23 \pm 0.78$ &$80.13 \pm 2.21$ &$71.83 \pm 2.19$ \\
\hline
ours (all)  &\bm{$6.90\pm0.35$} &\bm{$82.47\pm0.73$} &\bm{$74.00\pm0.57$} \\
\hline
\end{tabular}
\vspace{-4mm}
\end{center}
\end{table}

\subsection{Visualization}

\begin{figure}[t]

\centering
\includegraphics[width=0.48\textwidth]{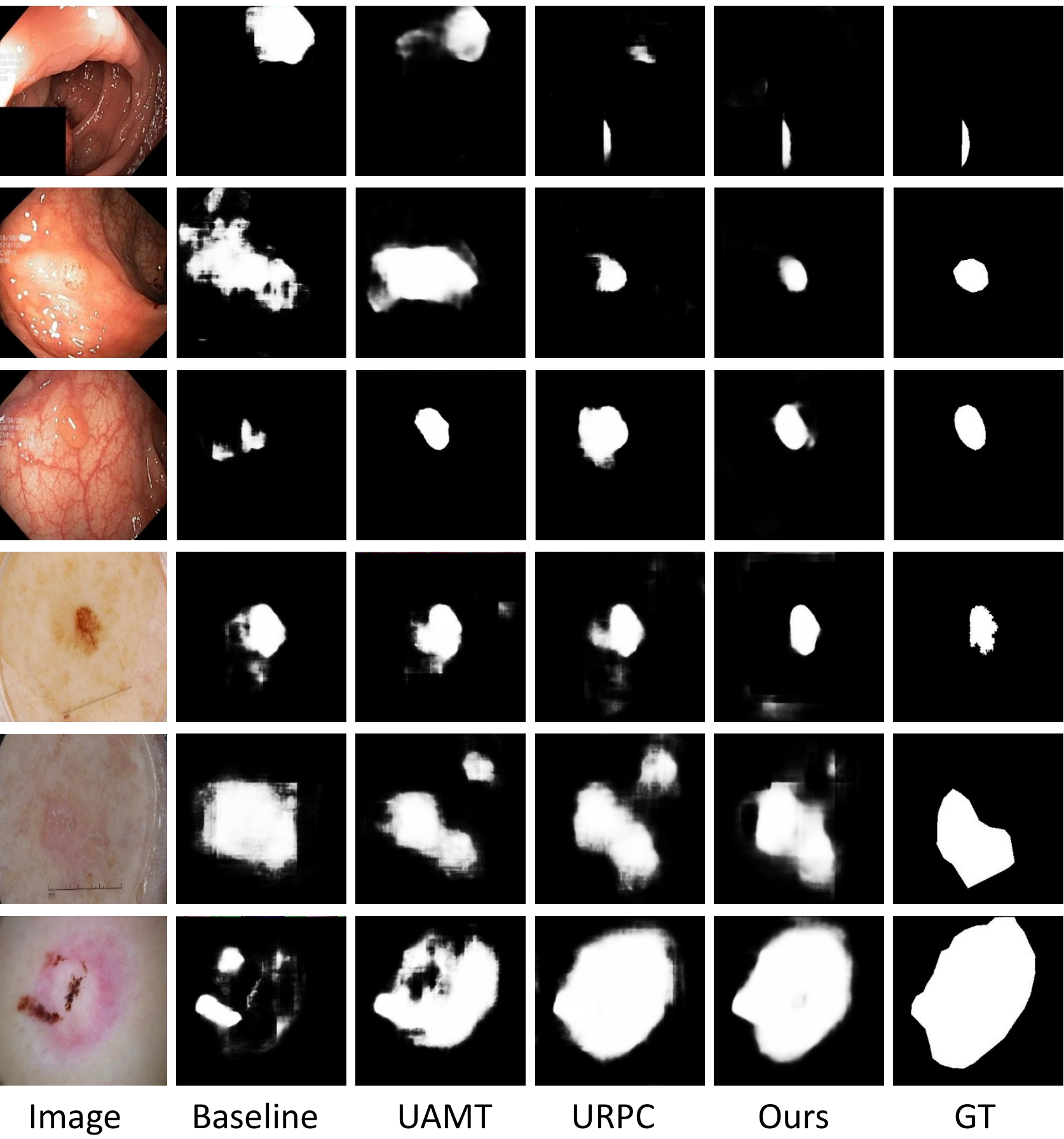}
\caption{Qualitative results for different approaches. Images in the upper three columns are from the Kvasir-SEG dataset, and images in the lower three columns are from the ISIC 2018 dataset.}
\label{fig:res}

\end{figure}

Figure \ref{fig:res} shows the visual segmentation results of our method on images of two datasets.
Three samples are presented for each dataset. 
It can be concluded from the images that existing methods can effectively process images with simple backgrounds and conspicuous objects 
However, these methods cannot fully segment all polyp or skin lesion regions when the object and background look similar,
or suppress all background regions when the background is cluttered.
In contrast, our approach can robustly better segment out object regions from background, even in complex cases.

\section{Conclusion}
In this work, we investigate the problem of semi-supervised medical image segmentation to reduce the human workload of labelling medical image data. 
We propose a novel semi-supervised segmentation approach by introducing a cross-level contrastive learning scheme and a cross-level consistency constraint. 
Our framework can use 
unlabelled data by exploring the intrinsic relationships across levels within the image. 
Extensive experiments on two public medical datasets show that our approach outperforms state-of-the-art semi-supervised learning methods consistently.

\bibliographystyle{IEEEbib}
\bibliography{strings,refs}

\begin{thebibliography}{10}

\bibitem{lee2013pseudo}
Dong-Hyun Lee et~al.,
\newblock ``Pseudo-label: The simple and efficient semi-supervised learning
  method for deep neural networks,''
\newblock in {\em Workshop on challenges in representation learning, ICML},
  2013, vol.~3, p. 896.

\bibitem{doersch2015unsupervised}
Carl Doersch, Abhinav Gupta, and Alexei~A Efros,
\newblock ``Unsupervised visual representation learning by context
  prediction,''
\newblock in {\em Proceedings of the IEEE international conference on computer
  vision}, 2015, pp. 1422--1430.

\bibitem{gidaris2018unsupervised}
Spyros Gidaris, Praveer Singh, and Nikos Komodakis,
\newblock ``Unsupervised representation learning by predicting image
  rotations,''
\newblock {\em arXiv preprint arXiv:1803.07728}, 2018.

\bibitem{hadsell2006dimensionality}
Raia Hadsell, Sumit Chopra, and Yann LeCun,
\newblock ``Dimensionality reduction by learning an invariant mapping,''
\newblock in {\em 2006 IEEE Computer Society Conference on Computer Vision and
  Pattern Recognition (CVPR'06)}. IEEE, 2006, vol.~2, pp. 1735--1742.

\bibitem{chen2020simple}
Ting Chen, Simon Kornblith, Mohammad Norouzi, and Geoffrey Hinton,
\newblock ``A simple framework for contrastive learning of visual
  representations,''
\newblock in {\em International conference on machine learning}. PMLR, 2020,
  pp. 1597--1607.

\bibitem{he2020momentum}
Kaiming He, Haoqi Fan, Yuxin Wu, Saining Xie, and Ross Girshick,
\newblock ``Momentum contrast for unsupervised visual representation
  learning,''
\newblock in {\em Proceedings of the IEEE/CVF Conference on Computer Vision and
  Pattern Recognition}, 2020, pp. 9729--9738.

\bibitem{xie2021propagate}
Zhenda Xie, Yutong Lin, Zheng Zhang, Yue Cao, Stephen Lin, and Han Hu,
\newblock ``Propagate yourself: Exploring pixel-level consistency for
  unsupervised visual representation learning,''
\newblock in {\em Proceedings of the IEEE/CVF Conference on Computer Vision and
  Pattern Recognition}, 2021, pp. 16684--16693.

\bibitem{lai2021semi}
Xin Lai, Zhuotao Tian, Li~Jiang, Shu Liu, Hengshuang Zhao, Liwei Wang, and
  Jiaya Jia,
\newblock ``Semi-supervised semantic segmentation with directional
  context-aware consistency,''
\newblock in {\em Proceedings of the IEEE/CVF Conference on Computer Vision and
  Pattern Recognition}, 2021, pp. 1205--1214.

\bibitem{wang2021dense}
Xinlong Wang, Rufeng Zhang, Chunhua Shen, Tao Kong, and Lei Li,
\newblock ``Dense contrastive learning for self-supervised visual
  pre-training,''
\newblock in {\em Proceedings of the IEEE/CVF Conference on Computer Vision and
  Pattern Recognition}, 2021, pp. 3024--3033.

\bibitem{chaitanya2020contrastive}
Krishna Chaitanya, Ertunc Erdil, Neerav Karani, and Ender Konukoglu,
\newblock ``Contrastive learning of global and local features for medical image
  segmentation with limited annotations,''
\newblock {\em arXiv preprint arXiv:2006.10511}, 2020.

\bibitem{ronneberger2015u}
Olaf Ronneberger, Philipp Fischer, and Thomas Brox,
\newblock ``U-net: Convolutional networks for biomedical image segmentation,''
\newblock in {\em International Conference on Medical image computing and
  computer-assisted intervention}. Springer, 2015, pp. 234--241.

\bibitem{jha2020kvasir}
Debesh Jha, Pia~H Smedsrud, Michael~A Riegler, P{\aa}l Halvorsen, Thomas
  de~Lange, Dag Johansen, and H{\aa}vard~D Johansen,
\newblock ``Kvasir-seg: A segmented polyp dataset,''
\newblock in {\em International Conference on Multimedia Modeling}. Springer,
  2020, pp. 451--462.

\bibitem{codella2019skin}
Noel Codella, Veronica Rotemberg, Philipp Tschandl, M~Emre Celebi, Stephen
  Dusza, David Gutman, Brian Helba, Aadi Kalloo, Konstantinos Liopyris, Michael
  Marchetti, et~al.,
\newblock ``Skin lesion analysis toward melanoma detection 2018: A challenge
  hosted by the international skin imaging collaboration (isic),''
\newblock {\em arXiv preprint arXiv:1902.03368}, 2019.

\bibitem{yu2019uncertainty}
Lequan Yu, Shujun Wang, Xiaomeng Li, Chi-Wing Fu, and Pheng-Ann Heng,
\newblock ``Uncertainty-aware self-ensembling model for semi-supervised 3d left
  atrium segmentation,''
\newblock in {\em International Conference on Medical Image Computing and
  Computer-Assisted Intervention}. Springer, 2019, pp. 605--613.

\bibitem{luo2020efficient}
Xiangde Luo, Wenjun Liao, Jieneng Chen, Tao Song, Yinan Chen, Shichuan Zhang,
  Nianyong Chen, Guotai Wang, and Shaoting Zhang,
\newblock ``Efficient semi-supervised gross target volume of nasopharyngeal
  carcinoma segmentation via uncertainty rectified pyramid consistency,''
\newblock {\em arXiv preprint arXiv:2012.07042}, 2020.

\bibitem{loshchilov2017decoupled}
Ilya Loshchilov and Frank Hutter,
\newblock ``Decoupled weight decay regularization,''
\newblock {\em arXiv preprint arXiv:1711.05101}, 2017.

\end{thebibliography}

\section*{Acknowledgments}
This work is supported in part by the Guangdong Basic and Applied Basic Research Foundation under Grant 
No. 
2020B1515020048, in part by the National Natural Science Foundation of China under Grant No.61976250 and No.62003256, in part by the Guangzhou Science and Technology Project under Grant 202102020633, and in part by the Guangdong Provincial Key Laboratory of Big Data Computing, The Chinese University of Hong Kong, Shenzhen.

\section*{Compliance with Ethical Standards}
This is a numerical simulation study for which no ethical approval was required.

\end{document}